\documentclass{article}

\usepackage{arxiv}

\usepackage[utf8]{inputenc} 
\usepackage[T1]{fontenc}    
\usepackage{hyperref}       
\usepackage{url}            
\usepackage{booktabs}       
\usepackage{amsfonts}       
\usepackage{nicefrac}       
\usepackage{microtype}      
\usepackage{lipsum}

\usepackage{graphics} 
\usepackage{epsfig} 
\usepackage{times} 
\usepackage{amsmath} 
\usepackage{amssymb}  
\usepackage{booktabs}
\usepackage[dvipsnames]{xcolor}
\usepackage{siunitx}
\usepackage{amsfonts}

\newtheorem{problem}{Problem}
\newtheorem{theorem}{Theorem}
\newtheorem{remark}{Remark}
\newenvironment{proof}{\paragraph{Proof:}}{\hfill$\square$}
\newtheorem{assumption}{Assumption}
\newtheorem{lemma}{Lemma}
\DeclareMathAlphabet{\mathbbold}{U}{bbold}{m}{n}
\usepackage{enumitem}   
\usepackage{algorithm}
\usepackage[noend]{algpseudocode}
\algrenewcommand\algorithmicrequire{\textbf{Input:}}
\algrenewcommand\algorithmicensure{\textbf{Output:}}

\usepackage{subcaption}
\usepackage{graphicx}
\usepackage{textcomp}
\usepackage{xcolor}
\usepackage{svg}
\usepackage{bbm}

\newcommand{\X}{{\mathsf{X}}}

\algdef{SE}[SUBALG]{Indent}{EndIndent}{}{\algorithmicend\ }%
\algtext*{Indent}
\algtext*{EndIndent}

\title{\LARGE \bf
Neural Network-based Co-design of Output-Feedback Control Barrier Function and Observer with Input Constraints
\thanks{This work was supported in part by ARTPARK and Siemens fellowship.}
}

\author{
 Vaishnavi Jagabathula $^\dag$\\
  Centre for Cyber-Physical Systems\\
  IISc, Bengaluru, India\\
  \texttt{vaishnavij@iisc.ac.in}\thanks{Authors contributed equally.} \\
   \And
 Ahan Basu $^\dag$\\
  Centre for Cyber-Physical Systems\\
  IISc, Bengaluru, India\\
  \texttt{ahanbasu@iisc.ac.in} \\
  \And
 Pushpak Jagtap \\
  Centre for Cyber-Physical Systems\\
  IISc, Bengaluru, India\\
  \texttt{pushpak@iisc.ac.in} \\
  }

\begin{document}
\maketitle
\thispagestyle{empty}
\pagestyle{empty}
\begin{abstract}
Control Barrier Functions (CBFs) provide a powerful framework for ensuring safety in dynamical systems. However, their application typically relies on full state information, which is often violated in real-world due to the availability of partial state information. In this work, we propose a neural network-based framework for the co-design of a safety controller, observer, and CBF for partially observed continuous-time systems with input constraints. By formulating barrier conditions over an augmented state space, our approach ensures safety without requiring bounded estimation errors or handcrafted barrier functions. All components are jointly trained by formulating appropriate loss functions, and we introduce a validity condition to provide formal safety guarantees beyond the training data. Finally, we demonstrate the effectiveness of the proposed approach through several case studies.
\end{abstract}
\section{Introduction}
Ensuring safety is crucial for deploying dynamical systems in real-world applications. Control Barrier Functions (CBF) \cite{ames2019control,cbf_stochastic} provide an efficient framework by encoding safety constraints as barrier functions, enabling controller synthesis that guarantees forward invariance of a safe set, typically via quadratic programming \cite{ames2019control}. The CBF framework has been extended to discrete-time \cite{dt_cbf} and stochastic systems \cite{clark2019control, cbf_stochastic}, through appropriate modifications to the safety conditions. Furthermore, CBFs have also demonstrated practical effectiveness in several applications \cite{ames2019control,sundarsingh2023scalable}. CBFs are often combined with techniques like model predictive control (MPC) \cite{zeng2021safety}, reinforcement learning (RL) \cite{cheng2019end}, or adaptive control \cite{taylor2020adaptive} to balance safety with performance. Despite its wide applicability, a major limitation of traditional CBF-based methods lies in their reliance on full state information. In realistic scenarios, such assumptions are prone to be violated due to inherently unobserved states.  Consequently, applying CBFs directly to partially observed systems can lead to performance degradation or even safety violations. 

Designing safety controllers for systems with unobserved or partially observed states remains a fundamental challenge in control theory. Prior works have approximated safe actions via supervised learning under bounded estimation error assumptions \cite{dean2021guaranteeing}, or required known error bounds \cite{cbf_obs_2, cbf_obs_3}. Another interesting approach \cite{jahanshahi2020synthesis} developed the barrier conditions for an augmented system of actual and estimated dynamics. However, such methods rely on the estimation error bounds, which are often difficult to guarantee in practice, and accurate observers for highly nonlinear systems are rarely feasible. In contrast, the construction of neural CBF has opened a gateway in synthesizing the safety controller, bypassing the need for hand-crafted barrier templates. Neural network based \cite{anand2023formally} and data-driven approaches \cite{Data_CBF_Zamani, jahanshahi2023synthesis} to construct CBF for unknown or partially observed systems have gained significant attention, but still assume either full-state access or bounded estimation error.

\begin{figure}
    \centering
    \includegraphics[width=0.9\linewidth]{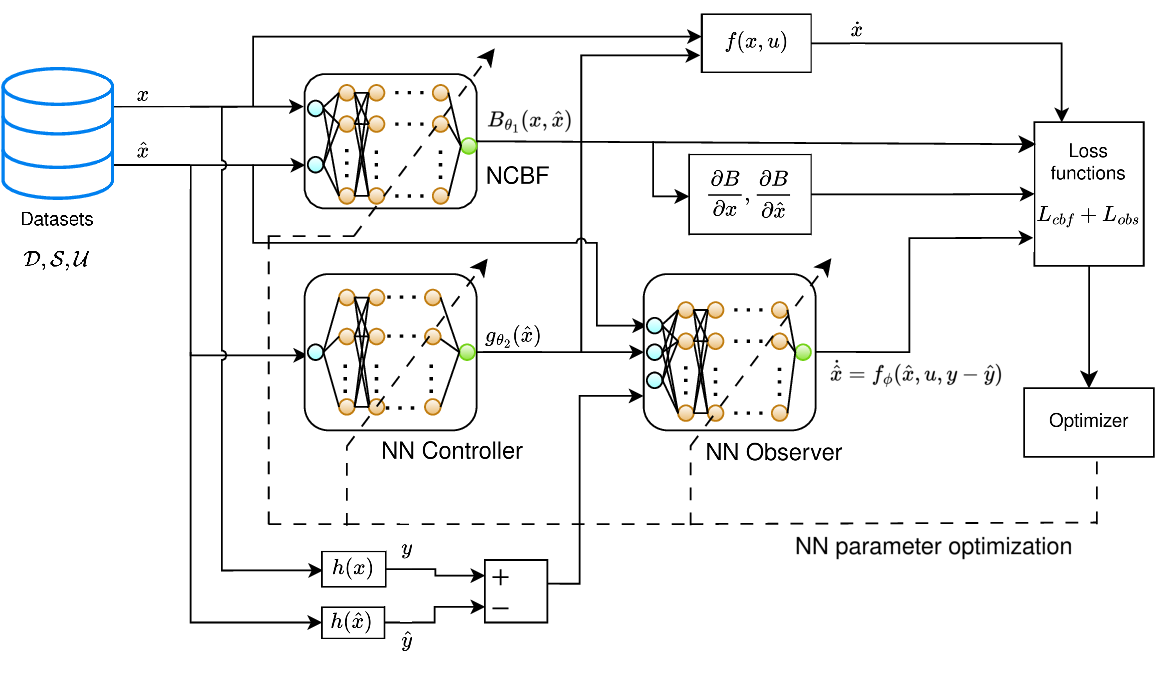}
    \caption{Simultaneous training of neural networks}
    \label{NN_training2}
\end{figure}

In this letter, we propose a neural network-based algorithm for the co-synthesis of a safety controller with observer and CBF for partially observed continuous-time systems under input constraints. {To the best of our knowledge, this is the first work that co-learns the controller, observer, and control barrier function concurrently for general nonlinear partially-observed systems and provides Lipschitz-based certification on-the-fly from sampled training data for continuous state space.} The key idea is to formulate CBF conditions over an augmented state space that satisfy the safety of the latent system. Our co-design strategy explicitly couples the controller with the observer’s estimates, yielding a cohesive synthesis, while the CBF formulation itself remains independent of the observer error, avoiding the need for an exact observer. To implement this framework, we present all three components as neural networks and train them simultaneously to ensure the safety guarantee of the actual system as seen in Figure \ref{NN_training2}. Though recent works have explored post-training verification of neural barrier functions using convex relaxations [\cite{mathiesen2022safety}] or SMT/CEGIS-based approaches [\cite{edwards2024fossil}], our approach integrates verification directly into the training loop by leveraging Lipschitz constants. This enables co-design of the neural CBF, controller, and observer that guarantee safety and enforce input constraints by construction, eliminating need of separate post-hoc verification. The framework’s effectiveness is demonstrated on partially observed (non)linear systems.

\section{Preliminaries and problem statement}
\textit{Notations}: 
The symbols $\mathbb{R}, \mathbb{R}_{\geq 0}$ denote the set of real and non-negative real numbers. We denote by $[1;N] := \{1,2,\ldots,N\}$ the set of natural numbers from $1$ to $N$. A column vector with $n$ rows of real number entries $x_1,..., x_n$ is denoted as $x=[x_1,...,x_n]^\top$, and the $n$-dimensional vector space is represented by $\mathbb{R}^n$. The symbol $\preceq$ denotes element-wise inequality of vectors. A vector space of real matrices with $n$ rows and $m$ columns is represented by $\mathbb{R}^{n\times m}$. A continuous function $\alpha:(-a,b) \mapsto \mathbb{R}$ for some $a,b>0$ is said to be an extended class $\mathcal{K}$ function if it is strictly increasing and $\alpha(0)=0$ and is denoted as $\mathcal{K}_e$. The partial differentiation of a function $f:X\times \hat{X}\mapsto \mathbb{R}$ with respect to the variable $x\in X$ is denoted by $\frac{\partial f}{\partial x}$. The Euclidean norm is represented using $||\cdot||$. The indicator function is defined as $\mathbbold{1}_{x\in X} =1$, if $x\in X$, $0$ otherwise. A function $f$ is Lipschitz continuous with Lipschitz constant $L$ if $||f(x_1)-f(x_2)||\leq L||x_1-x_2||$. 

\subsection{System Description and Problem Formulation}
Consider the continuous-time control system $S$:
\begin{align}
    S:=\big\{
    \dot{x}= f(x,u), \quad
    y = h(x) 
    \big\},\label{dyn}
\end{align}
where $x\in D\subset \mathbb{R}^n$, $u\in U\subset \mathbb{R}^m$, and $y\in Y\subset \mathbb{R}^p$ are the state, input, and output of the system, respectively. We assume that state space $D$, input space $U$ (capturing input constraints), and output space $Y$ are compact sets. 
The dynamics $f: D \times U\mapsto \mathbb{R}^n$ is assumed Lipschitz continuous in $x$ and $u$ over $D$ and $U$, with constants $L_x, L_u$, respectively and the output map $h: D \mapsto Y$ is Lipschitz continuous with constant $L_h$. Define $X:=\{x\in D \mid \inf\{\|x-y\|\mid y\in \partial D\}\geq\rho \}\subset D$, $\rho>0$, where $\partial D$ is the boundary of $D$.

In this letter, we consider partially observed systems where the full-state measurements of the system are unavailable (i.e., $p<n$) {but the system is fully controllable and observable}. Now we define the main problem of the paper.

\begin{problem}
    Given a partially observed continuous-time system $S$ as in \eqref{dyn} with input set $U$, an initial region $X_0\subset X\subset D$, and an unsafe region $X_u\subset X$, such that $X_0\cap X_u=\emptyset$, our objective is to design an output-feedback controller {(if exists)} for the system to ensure that the system trajectory starting from $x(0)\in X_0$ never enters the unsafe set $X_u$ {while remaining in $X$}, i.e., {$x(t)\in X \setminus X_u, \forall t\geq 0$}.
\end{problem}

For a partially observed system, in order to design a controller to ensure safety, we need to estimate the states using an observer. Inspired by structures of observers in \cite[Chapter 14]{khalil2002nonlinear}, \cite{zeitz1987extended}, we assume that the observer for $S$ can be represented as:
\begin{align}
    \hat{S}:=\big\{
    \dot{\hat{x}} = \hat{f}(\hat{x},u, y-\hat{y}),\quad
    \hat{y} = h(\hat{x})
    \big\},\label{dyn_obs}
\end{align}
{where $\hat{f}$ denotes the observer dynamical function with $\hat{x}, \hat{y}$ being the state and output of the observer, respectively.}
Next, we define the augmented state $\tilde{x} = [x^\top, \hat{x}^\top]^\top$, where $x$ and $\hat{x}$ are the states of $S$ and $\hat{S}$, respectively. The corresponding augmented system $\tilde{S}$ is defined as:
\begin{align}\dot{\tilde{x}}=
    \begin{bmatrix}
        \dot{x}\\\dot{\hat{x}}
    \end{bmatrix}=\begin{bmatrix}
        f(x, u)\\ \hat{f}(\hat{x},u, h(x)-h(\hat{x}))
    \end{bmatrix}.
\end{align}

\subsection{Control Barrier Function (CBF) for Augmented System}
We adopt a control barrier function (CBF) approach to design a controller that keeps system \eqref{dyn} away from unsafe regions. This section presents sufficient CBF-based conditions for the augmented system (following \cite{jahanshahi2020synthesis}) that guarantee safety of the original continuous-time system \eqref{dyn} starting from any initial state in $X_0\subset X{\subset D}$, remains in $X$, while avoiding the unsafe region $X_u\subset X$. {The overall unsafe region can be represented as $\overline{X}_u=X_u\cup (D\setminus X)$. We denote augmented unsafe region as $ \widetilde{X}_u:=(D\times D)\setminus (X\setminus{X_u}\times X)$}

Suppose there exists a continuously differentiable function $B:D\times D\mapsto \mathbb{R}$ and a continuous function $g:D\mapsto U$ satisfying:
\begin{align}
        \forall& (x, \hat{x})\in X_0\times X_0, B(x,\hat{x}) \geq 0, \label{cbf_1}\\
        \forall& (x, \hat{x})\in \widetilde{X}_u, B(x,\hat{x}) < 0 ,\label{cbf_2}\\
        \forall &(x, \hat{x})\in {D\times D}, \text{ s.t. } \frac{\partial B}{\partial x}f\big(x,g(\hat{x})\big)+ \frac{\partial B}{\partial \hat{x}}\hat{f}\big(\hat{x},g(\hat{x}), h(x)-h(\hat{x})\big) \geq -\alpha(B(x,\hat{x})),\label{cbf_3}
    \end{align}
where $\alpha$ is a class $\mathcal{K}_e$ function. Then we say that the function $B$ is a valid control barrier function constructed over the augmented system $\tilde{S}$ if it satisfies conditions \eqref{cbf_1}-\eqref{cbf_3}.

\begin{theorem}\label{th:barrier}
    Consider the partially observed system $S$ as in \eqref{dyn}, its state observer $\hat{S}$ as in \eqref{dyn_obs}, and the augmented system $\tilde{S}$. The initial and unsafe sets are ${X_0,\X_u \subset D}$. {Suppose there exists a continuously differentiable function $B$ under some controller $g$ that satisfies conditions \eqref{cbf_1}-\eqref{cbf_3}.} 
    Then, the system starting from any initial state $x_0\in X_0$ will stay outside the unsafe region, i.e., $x(t)\notin {\X_u}, \forall t\geq 0$.
\end{theorem}
\begin{proof}
   Since the system starts in $X_0$, as per \eqref{cbf_1} $B(x(0), \hat{x}(0))\geq 0$. Now, consider the condition \eqref{cbf_3}, there exists some control input $g(\hat{x})$ such that $\dot{B}(x,\hat{x}) \geq -\alpha(B(x,\hat{x}))$, where $\alpha(\cdot)$ is a class $\mathcal{K}_e$ function. Let $b(t) = B(x(t), \hat{x}(t))$. Then, $\dot{b}(t)\geq -\alpha(b(t))$. Let $\beta(t)$ be the solution to $\dot{\beta}(t) = -\alpha(\beta(t)), \beta(0) = b(0)$. Using the comparison lemma \cite[Chapter 3]{khalil2002nonlinear}, $b(t)\geq \beta(t), \forall t\geq 0$. Since $b(0)\geq 0,$ and $\beta(t)$ is non-negative for all $t$, we have $b(t)=B(x(t), \hat{x}(t))\geq \beta(t)\geq 0, \forall t\geq 0$. Therefore, {we have $(x(t),\hat x(t))\notin (D\times D)\setminus (X\setminus{X_u}\times X)$. Since $(x(t), \hat x(t))\in D\times D$, we can say $(x(t), \hat x(t))\in X\setminus X_u \times X$. This further implies that the actual and observer states are within their respective safe state space. $x(t)\in (X\setminus X_u)\implies x(t)\notin \X_u,$ and $ \hat x(t)\in X,\forall t\geq 0$.}
\end{proof}
\section{Formal verification of Neural Network based Observer and CBF}\label{section3}
In this section, we synthesize CBF satisfying \eqref{cbf_1}-\eqref{cbf_3} along with the state estimation-based controller for a partially observed continuous-time system. {We now provide the following lemma about the existence of the controller, barrier, and observer:
\begin{lemma}\label{lem:ROP}
    The partially observed system $S$ is assured to be safe if the following condition is satisfied with $\eta \leq 0$:
    \begin{align}
     \max(q_k(x,\hat{x}))\hspace{-0.2em} \leq\hspace{-0.1em} \eta, k\hspace{-0.2em}\in\hspace{-0.2em}[1;3],\label{ROP} \forall (x,\hat{x}) \hspace{-.2em}\in\hspace{-.2em} {D\hspace{-.2em}\times \hspace{-.2em}D},
     \end{align} 
     where 
     \begin{subequations}\label{eq:ROP}
         \begin{align}
         q_1(x,\hat{x}) &= -B(x,\hat{x}) \mathbbold{1}_{{(x,\hat{x})\in X_0\times X_0}}, \\
         q_2(x,\hat{x}) &= (B(x,\hat{x})+\delta)\mathbbold{1}_{{(x,\hat{x})\in \widetilde{X}_u}},\\
         q_3(x,\hat{x})
         &=-\frac{\partial B}{\partial x}f\big(x,g(\hat{x})\big) - \frac{\partial B}{\partial \hat{x}}\hat{f}\big(\hat{x},g(\hat{x}), h(x)-h(\hat{x})\big) -\alpha(B(x,\hat{x})), 
         \end{align}
     \end{subequations}
     with $\delta$ is a small positive value to enforce the strict inequality.
\end{lemma}
\begin{proof}
    The first condition with $\eta \leq 0$ ensures the positive semi-definite value of the barrier function within the initial region, while the second condition will ensure a strict negative value in the unsafe region due to the positive value of $\delta$. The third condition with $\eta \leq 0$ ensures the confinement of the system trajectory within the safe set as seen from Theorem \ref{th:barrier}. Therefore, satisfaction of condition \eqref{eq:ROP} with $\eta \leq 0$ ensures the system is safe under the controller $g$.
\end{proof}
}

{However, the main challenge of this framework is an infinite number of constraint satisfaction due to the continuous state-space. To overcome this, we use a finite number of samples from the augmented space $D \times D$. To do so, we collect $N$ samples $s^{(r)}:= (x,\hat x)^{(r)}$ from the set $D \times D$, where $r \in [1;N]$. Now, we consider a ball $\mathcal{S}_r$ around each sample $s_r$ with radius $\epsilon$, such that $\forall p \in D \times D$, there exists an $s^{(r)}$ satisfying
\begin{align}\label{eq:sample}
    \lVert p - s^{(r)} \rVert \leq \epsilon, \ \epsilon < \rho.
\end{align}
This ensures that the union of the balls forms a superset of the complete augmented space, i.e., $\bigcup_{r=1}^{N} \mathcal{S}_r \supset D \times D$. 
}

In addition, since some states are unmeasurable and the observer is unknown, we approximate the CBF, controller, and observer with neural networks $B_{\theta_1}, g_{\theta_2}, \hat{f}_\phi$, where $\theta_1, \theta_2, \phi$ are trainable parameters. Next, we provide the following assumption and lemma to state the main theorem of this section.

\begin{assumption}
    The candidate NCBF is assumed to be Lipschitz continuous with Lipschitz bound $L_b$, and its derivative is represented as a slightly modified NCBF neural network with changes reflected in the weights of the last layer and is Lipschitz bounded by a constant $L_{dB}$ [refer to \cite{basu2025neural} for more details]. The controller and observer neural networks have Lipschitz bounds $L_c$ and $L_o$, respectively. Additionally, the partial derivative $\frac{\partial B}{\partial x}$, functions $f(x,u)$, $\hat{f}_\phi$ and $h(x)$ are bounded by $M_{B}, M_f, M_o, M_h$ respectively, i.e., $\sup_x ||\frac{\partial B}{\partial x} || \leq M_B, \sup_{(x,u)} ||f(x,u)||\leq M_f$,  $\sup_{(\hat{x},u)} ||\hat{f}_\phi(\hat{x},u,y-\hat{y})||\leq M_o$, $|\sup_xh(x)|\leq M_h$. \label{assumption1}
\end{assumption}
\begin{lemma} \cite[Ex 3.3]{khalil2002nonlinear}
    If two functions $f$ and $g$ are Lipschitz continuous with constants $L_f$ and $L_g$, respectively, and are bounded by $\sup ||f||\leq M_f$ and $\sup||g||\leq M_g$, then their product $fg$ is also Lipschitz continuous with Lipschitz constant $M_fL_g+M_gL_f$. \label{lemma}
\end{lemma}
Now, we provide the main theorem of this section.
\begin{theorem}\label{th:verification}
    The partially observed system $S$ is guaranteed to be safe under the barrier, controller and observer $B_{\theta_1}, g_{\theta_2}, \hat{f}_\phi$ trained over the sampled points as in \eqref{eq:sample} if the following conditions are satisfied with $\hat{\eta} + L_{\max}\epsilon \leq 0$: 
    \begin{align}\label{eq:SOP}
        \max(q_k(s^{(r)}))\leq \hat{\eta}, k\in[1;3], \forall s^{(r)}\in D\times D,  \forall r\in[1;N],
    \end{align}
    with $q_1, q_2, q_3$ are as defined in \eqref{eq:ROP} and $L_{max} \hspace{-0.05cm}=\hspace{-0.05cm} \max\{L_b,\hspace{-0.05cm}M_f L_{dB} \\+ M_B(L_x+L_uL_c) + M_BL_o\sqrt{1+L_c^2+2M_h^2}+M_oL_{dB} +\alpha L_b \}$.
\end{theorem}
\begin{proof}
    This proof shows that if the condition $\hat{\eta} + L_{\max}\epsilon \leq 0$ is met with the satisfaction of the condition \eqref{eq:SOP}, then the system $S$ is safe. From \eqref{eq:sample}, for any $(x,\hat{x})\in D\times D$, $\exists (x,\hat x)^{(r)}\in D\times D$, s.t. $||(x,\hat{x})-(x,\hat{x})^{(r)}||\leq \epsilon$, and any $k\in\{1,2,3\}$,  $q_k(x, \hat{x}) = q_k(x, \hat{x})- q_k((x, \hat{x})^{(r)}) + q_k((x, \hat{x})^{(r)})\leq L_k||\tilde{x}-\tilde{x}^{(r)}||+\hat\eta\leq L_k\epsilon + \hat\eta\leq L_{max}\epsilon+\hat\eta \leq 0$, where $\tilde{x} = [x^\top, \hat{x}^\top]^\top$, Lipschitz constants of $q_k$ for $k\in[1;3]$ are computed using the Assumption \ref{assumption1} and Lemma \ref{lemma}, $L_1=L_2= L_b, L_3= M_fL_{dB}+M_B(L_x+L_uL_c) + M_BL_o\sqrt{1+L_c^2+2M_h^2}+M_oL_{dB}+\alpha L_b$. Therefore, satisfaction of the condition implies satisfaction of Lemma \ref{lem:ROP}, thereby guaranteeing the system to be safe.
\end{proof}

\begin{remark}
    {Note that the Lipschitz constant of the system can be estimated from physics-based models or data-driven methods \cite[Algorithm 2]{nejati2023formal}, while tighter Lipschitz bounds for the neural networks can be obtained using existing literature \cite{fazlyab2019efficient, pauli2024lipschitz}.}
\end{remark}
\begin{problem}
    For a partially observed continuous-time system \eqref{dyn} with initial set $X_0$ and unsafe sets $X_u$, our objective is to design an algorithm to synthesize an NCBF $B_{\theta_1}(x, \hat{x})$, a controller $g_{\theta_2}(\hat{x})$, and an observer $\hat{f}_{\phi}(\hat{x}, u, y-\hat{y})$, ensuring conditions in \eqref{eq:SOP} are satisfied. \label{problem_2}
\end{problem}

\section{Methodology}\label{Sec:Methodology}
In this section, we provide the methodology for solving Problem \ref{problem_2}, presenting the neural network architecture, the loss functions for the CBF constraints in \eqref{ROP}-\eqref{eq:ROP}, and the training process that ensures formal guarantees.

\subsection{Neural Network Architecture}\label{Sec:Architecture}
We denote the neural network architecture as $\{n^0, \{n^l\}^l,n_o\}$ with an input layer of $n^0$ neurons, $l$ number of hidden layers, each with $n^l$ neurons, and an output layer of $n_o$ neurons. The activation function applied to neurons is indicated by $\sigma(\cdot)$. Since we require the computation of $\frac{\partial B}{\partial x}$, a smooth activation function is preferred (for example, Softplus, Tanh, Sigmoid, etc.) for NCBF. 
The resulting neural network function is obtained by recursively applying the activation function in hidden layers. The output of each layer is given as $z_{k+1} = \Sigma_{k}(w_kz_k+b_k), \forall k\in\{0,1,..., l-1\}$, where $\Sigma_i(z_i) = [\sigma(z_{i}^{1}),...,\sigma(z_{i}^{n^i})]$ with $z_i$ as concatenation of outputs $z_i^j, j\in [1;n^i]$ of the neurons in $i$-th layer. For the $i$-th layer, weight matrices and bias vectors are denoted by $w_i\in\mathbb{R}^{n^{i+1}\times n^i}$ and $b_i\in \mathbb{R}^{n^{i+1}}$, respectively. For the NCBF and observer neural network, the final output is $y_{NN}(z_l) = w_lz_l+b_l$. {For the controller network, which must satisfy the input constraints $U:=\{u\in \mathbb{R}^m\mid \text{lb}\preceq u\preceq \text{ub}\}$, we bound the output between `lb' and `ub' using the HardTanh activation function as $y_{NN}(z_l)=\text{HardTanh}(w_{l}z_{l} + b_{l})_{\text{lb}}^{\text{ub}}$. The output of $\text{HardTanh}(x)_{\text{lb}}^{\text{ub}}=\text{lb} \text{ if }x<\text{lb}$, $\text{ub} \text{ if }x>\text{ub}$, $x$, otherwise.} 

The overall trainable parameter of a neural network is denoted by $\theta=[w_0, b_0,..., w_{l}, b_{l}]$.
Given an $n-$dimensional system with $m$ inputs and $p$ observations, the network architecture is as follows: NCBF $B_{\theta_1}: \{2n, \{n^l\}^l,1\}$, controller $g_{\theta_2}:\{n, \{n^l\}^l,m\}$ and the observer $\hat{f}_\phi:\{n+m+p, \{n^l\}^l,n\}$ with trainable parameters $\theta_1, \theta_2,\phi$, respectively.
\subsection{Formulation of Loss Functions and Training procedure}
We present the loss functions for training the NCBF, controller, and observer, whose minimization enforces the conditions \eqref{eq:SOP}.
Consider the partially observed system $S$ of the form \eqref{dyn}, an initial set $X_0\subset X$, and an unsafe set $X_u\subset X$. {We sample the points from an augmented set $D\times D$ as described in Section \ref{section3}. Consider the following sets: $\mathcal{S}=\{s^{(r)}\mid s^{(r)}\in X_0\times X_0, r\in[1;N]\}, \mathcal{U}=\{s^{(r)}\mid s^{(r)}\in \widetilde{X}_u, r\in[1;N]\}, \mathcal{D}=\{s^{(r)}\mid s^{(r)}\in D\times D, r\in[1;N]\}$.}

All three neural networks $B_{\theta_1}, g_{\theta_2}$, and $\hat{f}_\phi $ are trained such that the following sub-loss functions are minimized:
\begin{align}
    &\mathcal{L}_1(\theta_1) = \sum_{s^{(r)} \in \mathcal{S}} ReLU\big( - B_{\theta_1}(s^{(r)}) - \hat{\eta} \big), \label{l1}\\
    &\mathcal{L}_2(\theta_1) = \sum_{s^{(t)} \in \mathcal{U}} ReLU\big(B_{\theta_1}(s^{(r)}) + \delta - \hat{\eta} \big), \label{l2} \\
    & \mathcal{L}_3 (\theta_1, \theta_2, \phi) = \sum_{s^{(r)} \in \mathcal{D}} ReLU\bigg(-\frac{\partial B_{\theta_1}}{\partial x^{(r)}}f(x^{(r)},g_{\theta_2}(\hat{x}^{(r)})) -  \frac{\partial B_{\theta_1}}{\partial \hat{x}^{(r)}}{\hat{f}_\phi\big(\hat{x}^{(r)}, g_{\theta_2}(\hat{x}^{(r)}), h(x^{(r)})-h(\hat{x}^{(r)})\big)}\hspace{-0.1cm}\nonumber\\
    &\quad \quad\quad\quad\quad\quad\quad\quad\quad\quad - \alpha(B_{\theta_1}(s^{(r)}))- \hat{\eta}\bigg),\label{l3}
\end{align}
where $ReLU(z) =\max (0,z)$, $\alpha(\cdot)$ is the class $\mathcal{K}_e$ function.
In addition to these sub-loss functions, we also define a loss function for the observer neural network based on the Lyapunov function defined over the estimation error, in order not to produce random values for the state estimation. 

Let us denote the estimation error as $e = x-\hat{x}$. The error dynamics is $\dot{e} = f(x,g_{\theta_2}(\hat{x}))-\hat{f}_\phi(\hat{x},g_{\theta_2}(\hat{x}),h(x)-h(\hat{x}))$.
Now, let a candidate continuous Lyapunov function $V:\mathbb{R}^n\mapsto \mathbb{R}_{\geq 0}$ be defined as $V(e)=\frac{1}{2}e^\top e $, with the following properties:
$\forall (x,\hat{x})\in D\times D$, $\dot{V}(e) = {e^\top} \big(f(x,g_{\theta_2}(\hat{x})) -\hat{f}_\phi(\hat{x},g_{\theta_2}(\hat{x}), h(x)-h(\hat{x}))\big) \leq -\beta V(e)$, where $\beta>0$.
The property is reformulated as the following loss function for the observer neural network.
\begin{align}
    &\mathcal{L}_{4}(\phi) = \sum_{(x, \hat{x})^{(r)}\in D\times D} {e^\top} \bigg(f(x^{(r)},g_{\theta_2}(\hat{x}^{(r)}))-\hat{f}_\phi\big(\hat{x}^{(r)},g_{\theta_2}(\hat{x}^{(r)}), h(x^{(r)})-h(\hat{x}^{(r)})\big)\hspace{-0.3em}\bigg)\hspace{-0.2em} +\hspace{-0.2em} \beta V(e).\label{l_obs}
\end{align}
The overall neural network loss is the weighted sum of the above sub-losses, computed as:
\begin{align}
    \mathcal{L}_{cbf}(\Gamma) &= k_1\mathcal{L}_1(\Gamma) + k_2\mathcal{L}_2(\Gamma) + k_3\mathcal{L}_3(\Gamma)
    \label{cbf_loss_fn_nn}\\
    \mathcal{L}_{obs} &= k_{4}\mathcal{L}_{4}(\phi), \label{obs_loss_fn}
\end{align}
where $k_1,k_2,k_3, k_4>0$ are the weights corresponding to the sub-loss functions and $\Gamma = [\theta_1, \theta_2, \phi]$ are the trainable parameters. 

{Now, Assumption \ref{assumption1} is crucial as the formal guarantee over the trained networks is ensured by Theorem \ref{th:verification}, which, in turn, depends on the Lipschitz bounds. To train the neural networks with Lipschitz bounds, we use the lemma adopted from \cite[Theorem 2]{fazlyab2019efficient} as used earlier in \cite[Lemma 4.1]{basu2025neural}. Therefore, to satisfy the LMIs corresponding to the Lipschitz constants of different entities, we formulate the loss function,
\begin{align}\label{eq:loss_ineq}
    \mathcal{L}_M(\Gamma, \Upsilon)& = -c_{l_1}\log\det(M_{L_B}(\theta_1,\Lambda)) - c_{l_2}\log\det(M_{L_{dB}}(\hat{\theta}_1,\hat{\Lambda})) -c_{l_3}\log\det(M_{L_c}(\theta_2,\Tilde{\Lambda})) \notag \\
   & \quad\quad-c_{l_4}\log\det(M_{L_o}(\phi,\Bar{\Lambda})), 
\end{align}
where $c_{l_1}, c_{l_2}, c_{l_3}, c_{l_4} >0$ are weights for sub-loss LMIs, $\Upsilon = [\Lambda, \hat{\Lambda}, \Tilde{\Lambda}, \Bar{\Lambda}]$ and $M_{L_B}(\theta_1,\Lambda), M_{L_{dB}}(\hat{\theta}_1,\hat{\Lambda})$, $M_{L_c}(\theta_2,\Tilde{\Lambda})$ and $ M_{L_o}(\phi,\Bar{\Lambda})$ are the matrices corresponding to the bounds $L_B, L_{dB}, L_c$, and $L_o$, respectively.}

{The training procedure is shown in Algorithm \ref{NN_training}.
\begin{algorithm}
\caption{NN Training}
\label{NN_training}
\begin{algorithmic}[1]
    \Require System $S$, sets $X,D,X_0,X_u, U$ and {data set: $s^{(r)}\in D\times D, r\in\{1,\ldots,N\}$ with discetization $\epsilon$}
    \Ensure $B_{\theta_1}, g_{\theta_2}, \hat{f}_\phi$
    \State Select NN hyperparameters ($l, n^l$, activation function, optimizer, scheduler), $k_{obs}, k_1, k_2, k_3$
    \State Initialize trainable parameters ($\theta_1, \theta_2, \phi$). Initialize $\hat{\eta}$ as $\hat{\eta} = - L_{\max}\epsilon $
    \For{$i\leq Epochs$ (Training starts here)}
        \State Create batches of training data from the dataset
        \State Find batch loss $\mathcal{L}=\mathcal{L}_{cbf}+\mathcal{L}_{obs}$ using \eqref{l1}-\eqref{obs_loss_fn}
        \State Use ADAM optimizer [\cite{kingma2014adam}] to update
        $\theta_{1}^i, \theta_{2}^i,\phi^i$
        \State \textbf{If} Theorem \ref{th:convergence} is satisfied $\rightarrow$ \textbf{return} $B_{\theta_1, g_{\theta_2}, \hat{f}_{\phi}}$.
    \EndFor
    \State print(`No suitable barrier and controller found')
\end{algorithmic}
\end{algorithm}}

Upon loss convergence according to Theorem \ref{th:convergence}, the trained NCBF-based NN controller $g_{\theta_2}$ can be deployed with the trained observer $\hat{f}_\phi$, guaranteeing system safety. 
{\begin{theorem}\label{th:convergence}
    Consider a partially observed continuous-time system \eqref{dyn} with initial set $X_0\subset X\subset D$, and unsafe set $X_u\subset X$. Suppose a neural CBF $B_{\theta_1}$, a controller network $g_{\theta_2}$, and an observer network $\hat{f}_\phi$ are trained using finite data samples collected as in \eqref{eq:sample}, such that the loss $\mathcal{L}_{cbf} = 0, \mathcal{L}_M(\Gamma, \Upsilon) \leq 0$ and $\mathcal{L}_{obs}$ is minimized. Then, for any trajectory starting at any initial state in $X_0$, the trained controller neural network $g_{\theta_2}$ keeps the system safe by virtue of Theorem \ref{th:barrier} and \ref{th:verification}.
\end{theorem}
\begin{proof}
The loss $\mathcal{L}_{cbf}=0$ implies that the conditions of Theorem \ref{th:verification} have been satisfied with some $\hat{\eta}$ satisfying the condition $\hat{\eta} + L_{\max}\epsilon \leq 0$. Hence, the controller $g_{\theta_2}$ is trained to ensure that the system is in the safe set following conditions \eqref{cbf_1}-\eqref{cbf_3}. Now, the loss $\mathcal{L}_M(\Gamma, \Upsilon) \leq 0$ implies the satisfaction of Assumption \ref{assumption1} with the predefined Lipschitz bounds. Therefore, the satisfaction of the theorem ensures the system is safe under the action of the controller $g_{\theta_2}$.
\end{proof}}
\begin{remark}
    Since we trained NCBF $B_{\theta_1}$ on the data set of the augmented system, we do not require that the observer loss $L_{obs}$ in \eqref{obs_loss_fn} converge to zero for synthesizing the controller $g_{\theta_2}$. Even with a nonzero $L_{obs}$, Theorem \eqref{th:convergence} holds. 
\end{remark}

\begin{remark}\label{remark:4}
    {
    In practice, due to model and data complexity, or numerical precision, the loss may not converge exactly to zero. Therefore, we assume a tolerance of the order $10^{-6}$ to $10^{-4}$ for $L_{cbf}$ during implementation. The theorem states the sufficient condition to provide validity of the trained neural networks to enforce safety of the partially-observed system, which can be readily observed in case studies in Section \ref{Sec:Simul}.
    }
\end{remark}
\begin{remark}
    {The algorithm lacks a general convergence guarantee, but strategies such as reducing the discretisation parameter $\epsilon$ \cite{zhao2021learning} or adjusting neural network hyperparameters (architecture, weights, learning rate) \cite{nn_lr} can improve convergence.}
\end{remark}
\section{Simulation Results}\label{Sec:Simul}
This section validates and compares our methodology using the DC motor, pendulum, and the three-tank system against existing techniques. The class $\mathcal{K}_e$ function is set as $\alpha(x)=0.1 x$. We ran the simulations using Python on a computer equipped with a NVIDIA GeForce RTX 2080 Ti GPU.\\
{\textit{A.} \textit{DC Motor}:} Consider the DC Motor dynamics \cite{obeidat2013real} as:
\begin{align*}
    \dot{x} &= \begin{bmatrix}
        \dot{x}_1\\
        \dot{x}_2 \end{bmatrix} =
        \begin{bmatrix}
            \frac{-R}{L}x_1 - \frac{K_{dc}}{L}x_2 + \frac{V_{in}}{L}\\
            \frac{K_{dc}}{J}x_1 - \frac{b}{J}x_2
        \end{bmatrix},
        \quad
        y=x_2,
\end{align*}
where $x_1$, $x_2$ are armature current and rotational speed, with only $x_2$ observed. The input voltage $V_{in}$ bounded in $[-1,1]V$. Motor parameters are $J=0.01 kg.m^2$, $b=0.1 N.m.s$, $K_{dc}=0.01 V/rad/sec$, $R=1 \Omega$, $L=0.5H$.
We consider the state space $D = [-0.125,0.125]\times[-0.525,0.525]$, $X = [-0.1,0.1]\times[-0.5,0.5]$, an initial set $X_0=[-0.03,0.03]\times[-0.2,0.2]$, and an unsafe set $\overline{X_u}=[-0.5,-0.05]\times[-0.5,-0.3]\cup [0.05,0.1]\times[0.3,0.5]\cup (D \setminus X)$. 

\begin{figure*}[pt]
    \centering
    \includegraphics[width=\linewidth]{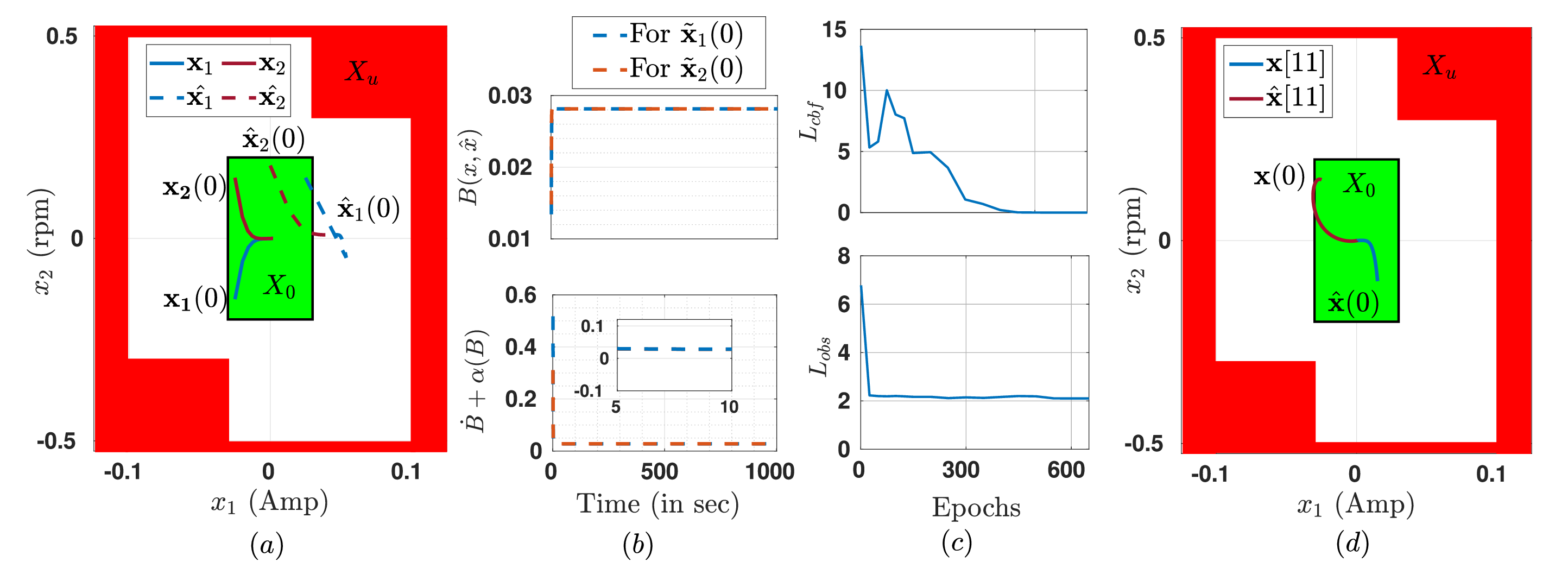}
    
    \caption{DC Motor (a) Actual and estimated trajectories starting from different initial conditions (b) Top: $B(x,\hat{x}) > 0$ for all $t \geq 0$ and Bottom: The satisfaction of condition \eqref{cbf_3} for different initial conditions, {(c) The CBF loss \eqref{cbf_loss_fn_nn} goes to zero, while the observer loss is minimized, (d) Controller in \cite{cbf_obs_3} ensures safety only if observer and actual state converges.}
    }
    \label{DC_motor}
\end{figure*}

{We fix the architecture of the three neural networks as NCBF: $\{4, \{64\}^4, 1\}$, NN observer: $\{4,\{128\}^5, 2\}$, and NN controller: $\{2, \{128\}^5, 1\}$.} We used Softplus activation function for hidden layers. The training algorithm converges to obtain neural networks for the observer $(\hat{f}_\phi)$, NCBF $(B_{\theta_1})$, and the controller $(g_{\theta_2})$ with $\hat\eta= -0.0532$ and discretization parameter $\epsilon = 0.023$. 
Using Algorithm \ref{NN_training}, with $\eta =- 0.0532<0$, loss convergence is satisfied, ensuring the safety. 
Figure \ref{DC_motor}(a) shows actual and estimated trajectories from different initial conditions in the augmented states $\mathbf{\tilde{x}_1}(0)=[\mathbf{{x}_1}(0)^\top, \mathbf{\hat{x}_1}(0)]^\top$ and $\mathbf{\tilde{x}_2}(0)=[\mathbf{{x}_2}(0)^\top, \mathbf{\hat{x}_2}(0)]^\top$. {Figure \ref{DC_motor}(b) shows that the trajectories satisfy the barrier conditions \eqref{cbf_1} and \eqref{cbf_3}, while Figure \ref{DC_motor}(c) illustrates that the CBF loss \eqref{cbf_loss_fn_nn} goes to zero and the observer loss \eqref{obs_loss_fn} is minimized to a finite value. Figure \ref{DC_motor}(d) highlights that, unlike traditional methods \cite{cbf_obs_3} requiring observer convergence, our approach maintains system safety even with an inaccurate observer (Figure \ref{DC_motor}(b)).}\\ 

\begin{figure*}[pt]
    \centering
    \includegraphics[width=\textwidth, ,height=0.325\linewidth,trim={0.22cm 6.5cm 1.0cm 1.0cm},clip]{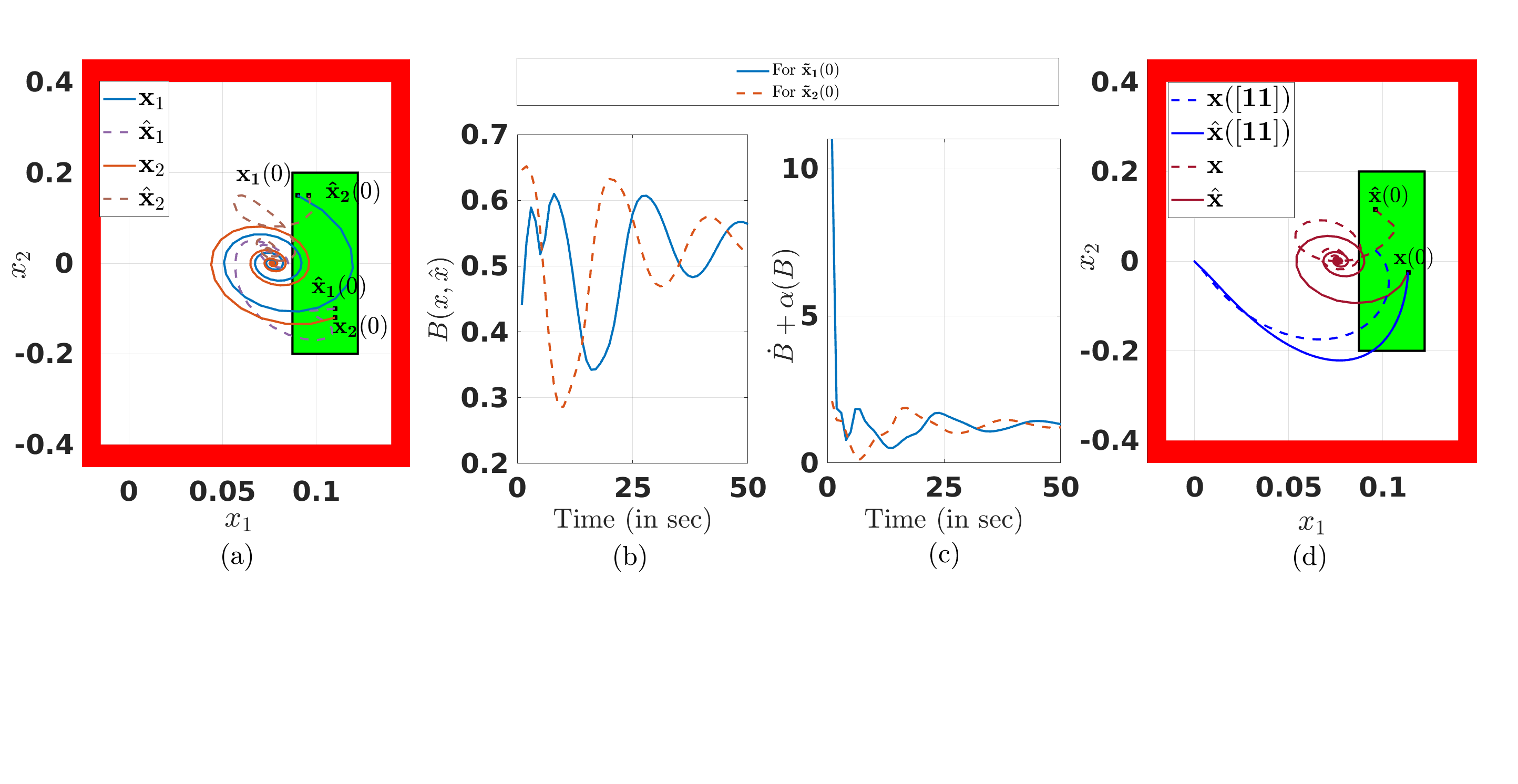}
    \caption{Pendulum (a) Actual and estimated trajectories starting from different initial conditions, (b)  The CBF $B(x,\hat{x})$ value is greater than zero for all time $t\geq 0$ for different initial conditions, (c) The condition \eqref{cbf_3} is satisfied for different initial conditions. {(d) Comparision of pendulum trajectories with \cite{cbf_obs_3}.}
    }
    \label{Pendulum}
\end{figure*}
\textit{B. Pendulum}:
We consider a simulation example of a pendulum with the following dynamics of the form \eqref{dyn}:
\begin{align*}
    \dot{x} &= \begin{bmatrix}
        \dot{x}_1\\
        \dot{x}_2 \end{bmatrix} 
        \begin{bmatrix}
            x_2\\
            \frac{u}{ml^2} - \frac{g}{l}\sin x_1 - \frac{b}{ml^2}x_2
        \end{bmatrix}, \ \
        y=x_1,
\end{align*}
where $x_1, x_2$ are the angle and angular velocity of the pendulum. The mass $m=1 kg$, the length of the rod $l=1$, damping coefficient constant $b=1$. $g=9.8 m/s^2$ is the acceleration due to gravity. The external input $u$ is torque, bounded within $[-1,1]$.
Let the state space $X =[-{\frac{2.7\pi}{180}}, \frac{10\pi}{180}]\times [-0.5, 0.5]$, initial region $X_0 = [\frac{5\pi}{180}, \frac{7\pi}{180}]\times [-0.3, 0.3]$, safe region $X_s = [-{\frac{\pi}{180}}, \frac{8\pi}{180}]\times [-0.4, 0.4]$, unsafe region $X_u =X\setminus X_s$. 
{We fix the architecture of the three neural networks as NCBF: $\{4, \{128\}^4, 1\}$, NN observer: $\{4,\{64\}^3, 2\}$, and NN controller: $\{2, \{32\}^2, 1\}$.}
We selected Softplus as the activation function for the hidden layers.
With $\hat\eta = -0.0538$ and discritization parameter $\epsilon=0.0193$, the training algorithm converges to obtain the neural networks for observer ($\hat{f}_\phi$), CBF ($B_{\theta_1}$) and controller $g_{\theta_2}$. 
Using Algorithm \ref{NN_training}, the loss convergence is satisfied, ensuring the safety of the system. The actual and estimated trajectories starting with different initial conditions in augmented space $\mathbf{\tilde{x}_1}(0)=[\mathbf{{x}_1}(0)^\top, \mathbf{\hat{x}_1}(0)]^\top$ and $\mathbf{\tilde{x}_2}(0)=[\mathbf{{x}_2}(0)^\top, \mathbf{\hat{x}_2}(0)]^\top$ are shown in Figure \ref{Pendulum} (a). 
From Figures \ref{Pendulum}(b) and \ref{Pendulum}(c), we can observe that the trajectories also satisfy the barrier conditions \eqref{cbf_1} and \eqref{cbf_3}, respectively. {We have used the trained controller on the actual system, which renders the system safe as shown in Figure \ref{Pendulum}(a).} {Detailed discussion of comparison is provided in Section \ref{Sec:Simul}.D.}

\begin{figure}
    \centering
    \includegraphics[width=0.8\linewidth]{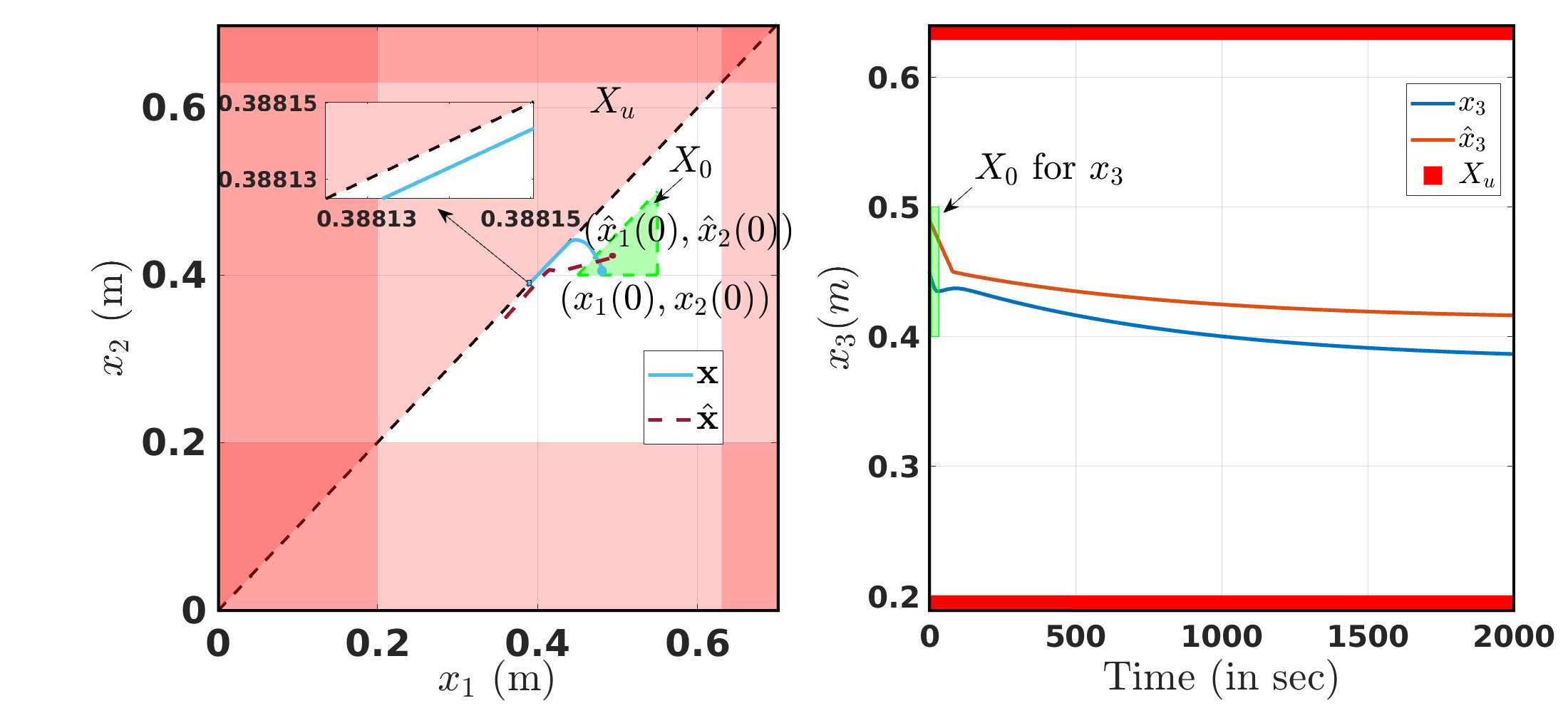}
    \caption{{Three Tank System: The trained controller ensures the states $x_1, x_2, x_3$ remain in the safe region.}}
    \label{fig:tank_level}
\end{figure}

\textit{C. Three Tank Level Control System:} 
{Consider a three-tank system of the following dynamics \cite{three_tank_citation}:
\begin{align}
    \dot{x}_1 &= -a_1 sgn(x_1-x_2)\sqrt{|x_1-x_2|}+u_1/A,\nonumber\\
    \dot{x}_2 &= a_1 sgn(x_1\hspace{-0.1cm}-\hspace{-0.1cm}x_2)\sqrt{|x_1-x_2|}\hspace{-0.1cm}-a_2sgn(x_2\hspace{-0.1cm}-\hspace{-0.1cm}x_3)\sqrt{|x_2-x_3}|,\nonumber\\
    \dot{x}_3 &= a_2 sgn(x_2-x_3)\sqrt{|x_2-x_3|}-a_3\sqrt{x_3}+u_2/A,\nonumber\\
    y&=[x_1,x_2]^\top,
\end{align}
}
{where $x_1,x_2,x_3$ are water levels in the tank, $u_1,u_2$ represent the total water inflows, and $sgn(\cdot)$ denotes the signum function. The coefficients $a_1,a_2,a_3$ and cross-section area of tanks $A$ are taken from \cite{three_tank_citation}. Let the state space be $D = [0, 0.7]^3$, $X=[0.1,0.63]^3$, and we aim to maintain the water levels in a safe region $X_s = \{(x_1,x_2,x_3)\in X\mid 0.2\leq x_2<x_1\leq 0.63, 0.2\leq x_3\leq 0.63\},$
 unsafe region $X_u=X\setminus X_s$, $\overline{X_u} = X_u\cup (D\setminus X) $, starting in $X_0=\{(x_1,x_2,x_3)\in X\mid 0.4\leq x_2\leq 0.5, x_2+0.05\leq x_1\leq 0.55, 0.4\leq x_3\leq 0.5\}$. Maximum allowable water inflow is $10^{-4}m^3/s$, i.e., $U=[0,10^{-4}]^2$. 
We fix the architecture of the three neural networks as NCBF: $\{6, \{128\}^4, 1\}$, NN observer: $\{7,\{64\}^3, 2\}$, and NN controller: $\{3, \{32\}^2, 2\}$.
We use 'Tanh' as the hidden-layer activation
With $\hat\eta = -0.0529$ and $\epsilon=0.0957$, the training algorithm converges to obtain the neural networks for observer ($\hat{f}_\phi$), CBF ($B_{\theta_1}$), and controller $g_{\theta_2}$. 
Since the loss converges, Theorem \ref{th:convergence} also holds, ensuring state trajectories remain safe, which is evident from Figure \ref{fig:tank_level}.\\
} 
\textit{D. Discussion:} 
We compare our approach for the pendulum and DC motor with \cite{cbf_obs_3} (Figures \ref{Pendulum}(d), \ref{DC_motor}(d)). In \cite{cbf_obs_3}, safety relies on the assumption of uniformly bounded estimation error, which
eventually decays with time \cite[Definition 5]{cbf_obs_3}. Constructing such accurate observers is difficult for highly nonlinear systems \cite{zeitz1987extended}, making the method restrictive. In contrast, our approach only assumes that the observer and the system start in the same initial region and does not require convergence to a zero observer loss to ensure safety (which is evident from Figure \ref{DC_motor}(a)). Moreover, existing works \cite{cbf_obs_1, cbf_obs_2, cbf_obs_3} often fix the CBF structure, which may fail under input constraints due to the feasibility of the QP-based formulation. Our method co-designs both the controller and barrier via training, explicitly incorporating input bounds (using HardTanh as discussed in Section \ref{Sec:Architecture}). The main limitation is scalability: training took ~2.5h for 2D systems and ~5.5h for 3D, suggesting exponential growth with dimension due to sampling in augmented state-space. Reducing this computational cost is an important direction for future work.
\section{Conclusion}
This study presents a framework for synthesizing a formally validated neural network controller that ensures safety in partially observed systems. Unlike conventional approaches, our design does not require the observer states to converge exactly to the true states. We reformulate CBF safety constraints into appropriate loss functions for finite state space data, and together with a validity condition, we provide safety guarantees for the continuous state space. We then propose a co-design training framework for jointly learning the controller and observer.

\bibliographystyle{IEEEtran}
\bibliography{references.bib}
\end{document}